\documentclass[]{article}
\usepackage[margin=1in]{geometry}

\usepackage{graphicx}
\usepackage{xcolor}
\usepackage{booktabs, multirow} % for borders and merged ranges
\usepackage{soul}% for underlines
\usepackage{changepage,threeparttable} % for wide tables

\usepackage{caption}

\usepackage{algorithmicx}
\usepackage{algorithm}% http://ctan.org/pkg/algorithms
\usepackage{algpseudocode}% http://ctan.org/pkg/algorithmicx

\usepackage{hyperref}

\begin{document}

\setlength{\parindent}{0pt}
\setlength{\parskip}{5pt}

\begin{center}
    \textbf{{\large Thinking beyond chatbots' threat to education: Visualizations to elucidate the writing and coding process}} \\
    \vspace{5pt}
    { Badri Adhikari} \\
    \vspace{4pt}
    {\small \textsuperscript{} \texttt{adhikarib@umsl.edu}} \\
    \vspace{3pt}
    {\small Department of Computer Science, University of Missouri-St. Louis, USA}
    \linebreak
\end{center}

% § 1. Why did you do this study or project?
% Problem Statement. Why is this problem important? Who cares?
% § 2. What did you do and how?
% Approach: Did you use simulation, analytic models, prototype construction, or analysis of field data for an actual product? What tool did you use to solve the problem? What was the extent of your investigation:  Did you look at one application or 100? Did you create surveys that are statistically valid?
% § 3. What did you find?
% Results: Did you find something that can improve speed, increase efficiency and/or reduce cost? Define your results in numbers.
% § 4. What do your findings mean?
% Conclusion: How will your work help your research community? What are the real-world applications? Establish the importance of your work, the difficulty of the area, and the impact it might have if successfully utilized.

\begin{abstract}
%Language is regularly credited with being the greatest achievement of humankind. It allows us to communicate intricate, complex ideas and (even in the simplest terms) deep emotions with very little effort.
The landscape of educational practices for teaching and learning languages has been predominantly centered around outcome-driven approaches. The recent accessibility of large language models has thoroughly disrupted these approaches. As we transform our language teaching and learning practices to account for this disruption, it is important to note that language learning plays a pivotal role in developing human intelligence. Writing and computer programming are two essential skills integral to our education systems. What and how we write shapes our thinking and sets us on the path of self-directed learning. While most educators understand that `process' and `product' are both important and inseparable, in most educational settings, providing constructive feedback on a learner's formative process is challenging. For instance, it is straightforward in computer programming to assess whether a learner-submitted code runs. However, evaluating the learner's creative process and providing meaningful feedback on the process can be challenging. To address this long-standing issue in education (and learning), this work presents a new set of visualization tools to summarize the inherent and taught capabilities of a learner's writing or programming process. These interactive \textit{Process Visualizations} (PVs) provide insightful, empowering, and personalized process-oriented feedback to the learners. The toolbox is ready to be tested by educators and learners and is publicly available at \href{https://www.processfeedback.org}{www.processfeedback.org}. Focusing on providing feedback on a learner's process---from self, peers, and educators---will facilitate learners' ability to acquire higher-order skills such as self-directed learning and metacognition.

\end{abstract}

\vspace{10pt}


\vspace{10pt}

\section{Introduction}

\subsection{Metacognitive writing and coding}

\textbf{Writing.} Writing shapes our thinking, regardless of whether we are writing to learn or learning to write \cite{langer1987writing, klein2016trends}. During the process, learners write in order to clarify their thinking \cite{whitney2008beyond}. Writing facilitates a logical and linear presentation of ideas that allows writers to explain their points of view. Emig \cite{emig1977writing}, a prominent 20\textsuperscript{th}-century English composition scholar, was best known for her foundational work relating to the process theory of composition, where writing is presented as a process rather than a product. Her work provided a fundamental way to develop new ideas by teaching students to develop their writing skills \cite{emig1977writing}. The effective teaching of writing is an essential component of any successful school program. Writing skills are best taught when related to specific and relevant content, which requires a more analytical approach to writing and provides a particularly welcoming context for thinking deeply. However, the recent use of online resources and chatbots reduces the motivation to write for users who let the chatbot write for them. The role of writing in the development of logical thinking is as important now as it has ever been.

\textbf{Coding.} Most schools and universities worldwide require students to learn computer programming. Several studies have pointed out that learning to code can improve various skills. In 1984, Pea and Kurland \cite{pea1984cognitive} identified a difference in learning programming compared with other studies. Their objective was to improve the \textit{pedagogy of programming}. They were among the first to teach computer programming, as they worked with high school students trying to determine the best approach to enabling students to see beyond the programming, so they could eventually acquire analogical, conditional, procedural, and temporal reasoning skills, as well as higher cognitive skills. They were looking for a \textit{cognitive style} that identified a student as being successful or not successful in performing specific programming subtasks. However, they never labeled any of the learners they analyzed as unable to learn to program. They emphasized teachers being trained to respond appropriately to a student's cognitive learning style. In a subsequent work, however, the idea was refuted \cite{kurland1986study}. A recent review revisiting the topic finds that learning to code can improve a learner's skills in mathematics, problem-solving, critical thinking, socialization, self-management, and academic strategizing \cite{popat2019learning}. 
%They found that only expert coders were conditioned to apply higher levels of deduction based on their years of experience. 
These findings suggest that the influence of learning to write a computer program on a learner's cognitive abilities can be anywhere from trivial to profound. It could simply be learning a vocabulary of commands and rules for arranging the commands. Or, it could be far more than programming, and acquiring powerfully general higher cognitive skills such as planning abilities.
%On this issue, the original work \cite{pea1984cognitive} had examined both of these extreme beliefs and found that learners with a low-level understanding of programming (i.e., ``grammatical" programming) may not gain higher-level cognitive skills. On the other hand, learners who can distance themselves sufficiently from the low-level coding aspects of program generation to reflect on the phases and processes of problem-solving (such as issues of elegance, optimization, efficiency, verification, style, etc.) can lead the learners to consider broader issues. 
Irrespective of the magnitude of impact, learning to build useful software can serve as a means for learners to attain and improve widely applicable cognitive skills.
%
%To what extent learning to code helps higher order thinking skills 
% They also  continued to put the success  of beginning coders in the hands of trained teachers who could coax higher levels of expectation under the right circumstances. 

\textbf{Learning metacognition.} 
Metacognition is commonly defined as the ability to think about one's own thinking. The text which inspired this definition was provided by John Flavell in 1979. He introduced it as the learners `knowledge of their own cognition' and formally defined metacognition as the `knowledge and cognition about cognitive phenomena' \cite{flavell1979metacognition, georghiades2004general}. Using metacognitive thinking and strategies enables learners to become flexible, creative, and self-directed. Metacognition assists learners with additional educational needs in understanding learning tasks, self-organizing, and regulating their own learning. The National Research Council's book, titled \textit{How People Learn: Brain, Mind, Experience, and School} \cite{national2000people}, stated:

\begin{quote}
``Metacognition also includes self-regulation—the ability to orchestrate one's learning: to plan, monitor success, and correct errors when appropriate—all necessary for effective intentional learning... Metacognition also refers to the ability to reflect on one's own performance." (National Research Council, 2000, p. 97).
\end{quote}

The most intriguing and difficult part of teaching metacognition is that thinking is idiosyncratic. Brame (2019) \cite{brame2019science}, however, tells us not to be afraid to ask questions and to work on our own self-directed, self-monitoring, and self-evaluating plans. Effective self-directed planning involves understanding the overall goal and timeline of a project at hand. Similarly, self-monitoring involves investigating why one failed to master a needed skill, and self-evaluating involves assessing why one was or was not successful and efficient. Overall, learners can develop higher-order skills when they ask themselves the questions needed to help them ``plan, monitor, and evaluate" their learning experiences.

\textbf{Metacognitive writing and coding.} Writing and coding can now be outsourced to generative deep learning systems. In the coming age, when we are increasingly outsourcing repetitive cognitive tasks to deep learning systems, it is crucial for learners to develop higher-order cognitive skills.
To learn to solve problems in authentic situations, learners should develop metacognitive abilities to manage and regulate their problem-solving process. For example, in the domain of mathematics, metacognition is a fundamental component of the problem-solving process \cite{hancock2021supporting}. Ironically, these cognitive tasks, such as writing and coding, are learners' vehicles to organize their thoughts and develop several higher-order thinking skills. Metacognition does not develop on its own for all learners; thus, it must be a target of learning supported explicitly in the classroom \cite{baten2017relevance}. Langer (1986) \cite{langer1986children} found that students at all grade levels were deficient in higher-order thinking skills. Today's outcome-oriented learning and education priorities make it even more challenging to develop such skills.
In general, developing metacognition and other related skills such as self-awareness, understanding our own mind, and critical thinking, require practicing reflection, mindfulness, learning from experience, engaging in self-experimentation, and seeking feedback. A potential solution to improving metacognition in learners is process-focused thinking aimed at developing process-oriented metacognition.

\subsection{Process-oriented feedback}

\textbf{Importance of process-focus.} Research on learning, directly or indirectly, embraces the ``process pedagogy." For instance, in direct opposition to the focus on written products, the 1970s and 1980s observed a groundswell of support for ``process" approaches to teaching writing \cite{langer1987writing}. In writing, learners and educators use words such as drafting, prewriting, and revising in commonplace speech \cite{whitney2008beyond}. Similarly, in learning computer programming, educators often require learners to submit process data such as code comments, challenges faced, and code optimizations as a part of the assignment requirements. Some programming approaches, such as `test-first' encourage programmers to write functional tests before the corresponding implementation code can be entered \cite{erdogmus2005effectiveness}.
%Unfortunately, despite sufficient evidence, educators, by default, focus on the product. 
If process-oriented teaching and learning are so essential, why are today's education systems mostly outcome-oriented? One key reason is that assessing the process and providing feedback is not straightforward.

\textbf{Importance of feedback.} Feedback is one of the most powerful influences on learning and achievement \cite{hattie2007power}. If effective, feedback can promote student investment and independence in writing and coding \cite{whitney2008beyond}. Detailed and descriptive feedback is found to be effective when given alone, unaccompanied by grades or praise. And surprisingly, the perceived source of feedback (a computer or an instructor) is found to have little impact on the results \cite{lipnevich2009effects}. Peer feedback can also be effective. For example, a study to examine the effects of peer-assessment skill training on learners' writing performance found that learners who receive in-depth peer assessment outperform those who do not \cite{xiao2010effects}. Similarly, self-assessment can have a powerful impact on learner motivation and achievement \cite{cauley2010formative}. Self-assessment is much more than simply checking answers. It is a process in which learners monitor and evaluate the nature of their thinking to identify strategies that improve understanding \cite{mcmillan2008student}. Within all the kinds of feedback that can be provided to learners, feedback on the process can be considerably more powerful.

\textbf{Importance of feedback on the process.} 
Providing feedback on a learner's thought process can have a more positive impact on learning than feedback focused on the final outcome \cite{paulson2017impact}. Across a wide range of educational settings, formative-type assessments and feedback do `work' and effectively promote student learning \cite{black1998assessment}. They provide valuable information to both learners and educators \cite{cauley2010formative}. Formative assessments, in general, are a tenet of good teaching. Unfortunately, approaches to assessment have remained inappropriately focused on testing \cite{shepard2000role}. Pressures on education are threatening the use of formative assessments \cite{yorke2003formative}. Teachers are, and continue to be under enormous pressure to get their learners to achieve \cite{dixson2016formative}. Moreover, in practice, these formative assessments make learners follow a prescribed and explicit process with \textit{a lot of steps}, potentially developing \textit{learned dependence} \cite{boud1995assessment}. In addition, the practice of \textit{lock stepping} learners by forcing them to divide their effort into stages conflicts with the recursive nature of the creative endeavor processes \cite{whitney2008beyond}. However, a more significant challenge is that providing feedback on the process instead of the product is difficult.

\subsection{Computational technologies to cultivate process feedback}

\textbf{Lack of computational technologies to provide feedback on the process.} Learners can be provided with process feedback in several areas of the creative process. Process activities in writing are often subdivided into stages such as prewriting, drafting, revising, and editing. These processes are recursive rather than linear and complex rather than simple \cite{langer1987writing}. To help learners \textit{prepare to write} or \textit{prepare to code}, educators can offer learners models of finished writing or code, or more effectively, can provide models on the spot, in front of learners, for a task in question. Models that demonstrate \textit{how} to plan and write are more effective than the ones that demonstrate \textit{what} to write \cite{whitney2008beyond}. In other words, in addition to providing professional examples, locally produced examples and examples written by other learners of similar age and skill level help learners imagine how to craft a product incorporating similar features \cite{whitney2008beyond}. However, effective tools to explore, analyze, and learn from the process that leads to the product, are missing. The crux of the problem lies in the absence of effective computational technologies for today's educators to demonstrate and discuss their own process of generating a product or to present and analyze a \textit{model} process and for today's learners to acquire feedback on their process.

\textbf{Overview of this work.} This article presents, discusses, and demonstrates how maintaining a self-monitored revision history (\textit{snapshots}) of one's own text as a writer or a programmer's code at frequent intervals during the writing or coding process, can be used to build interactive data visualizations that display and summarize the process. These Process Visualizations can provide insights into the steps a writer or a programmer took during their writing or coding process. This ongoing self-review by the author can facilitate effective self-feedback, peer feedback, and educator feedback.

\section{Methodology and Implementation}

This section describes the overall approach of retaining a user's process data (typed text of a writer or code or a programmer) and using that data to render informative data visualizations. First, the learner must realize that the web application being used was designed to allow users to write text or code. Once a user starts to type, the web application stores the text/code typed every five seconds in the user's local storage (Internet browser storage). These timestamped data contain the revision history of the user-typed text over the entire typing period. In other words, every entry recorded in this revision history contains both a timestamp and the text the user was working with during time stamp process. Together, the full revision data collection tool captures how the typed text/code evolves (or changes) over time.

\textbf{Connecting passages over versions.} Transforming typed text data (e.g., an English essay) with revision history information (i.e., time and full text at that time) into a format that can be used to develop visualizations requires that each text passage at any given time step can be connected to the text in the following time step. In this case, a passage is defined as either a sentence or a paragraph. These passage-to-passage connections between two versions of the text can then be used to observe the writer's creative flow when writing a passage, i.e., how a passage evolves (or changes) over time. For instance, as a writer adds content to a paragraph, the content of that paragraph will change with each timestep. However, the various versions of a paragraph must be identified as the same to correctly display the paragraph's origin (i.e., the time step when it was introduced). An algorithm that usefully maps passages in a text at time $t$ with passages in the next step (at time $t+1$) should also account for the fact that new passages can be added anywhere in the middle of the text at any time during the writing process. 
%In other words, mapping passages, in two text versions, based on the order in which they appear in the text would not work.

\textbf{Maintaining passage identities.} Identifying the same passages over multiple text revisions is a crucial step toward generating meaningful visualizations. The idea is to initially assign a random identity (ID) to each passage in each text revision, i.e., all passages are assigned different IDs regardless of their similarity. Next, these passage IDs are updated by calculating their similarities. Specifically, the update process starts from the most recent pair of two text versions. First, similarities between all passages at the time step $t+1$ and all those at time $t$ are calculated (i.e., pairwise matched). For the two passages $p$ and $q$, that have the highest similarity score, where $p$ and $q$ are passages at time $t+1$ and $t$ respectively, the ID of passage $p$ is updated to that of $q$ if the similarity is greater than a predefined similarity threshold. In other words, passage IDs are propagated from the most recent time step to previous time steps. This process is repeated for all adjacent time step pairs all the way to the first pair of revisions (see \textbf{Algorithm \ref{myalgorithm}}).

% Resource: https://shantoroy.com/latex/how-to-write-algorithm-in-latex/
%\renewcommand{\thealgorithm}{} % For removing numbering
\renewcommand{\algorithmicrequire}{\textbf{Inputs:}}
\renewcommand{\algorithmicensure}{\textbf{Outputs:}}
\algnewcommand\algorithmicforeach{\textbf{for each}}
\algdef{S}[FOR]{ForEach}[1]{\algorithmicforeach\ #1\ \algorithmicdo}
\begin{algorithm}
\small
\caption{Identify same passages (paragraphs or sentences) between two versions of text}
\label{myalgorithm}
\begin{algorithmic}[1]
\Require~~
\Statex {$S$, list of text data at various time steps; each item consists of an ordered sequence of passages with IDs}
\Statex {$threshold$, similarity threshold (between 0 and 1)}
\Ensure~~
\Statex {$S'$, IDs of passages at time $t_{n}$ replaced with matching passages at time $t_{n+1}$}
\Procedure{UpdateIDs}{}
\ForEach {pair of timestamps $t$ and $t+1$, starting from the latest pair}
    \ForEach {$m \in S_{t} $}   \Comment{$m$ is a passage at time step $t$}
        \State $n_{max}$ $\gets$ $-1$ \Comment{Stores the similarity of the most similar passage in $t+1$}
        \State $n_{id}$ $\gets$ $null$ \Comment{Stores the ID of the corresponding passage}
        \ForEach {$n \in S_{t+1} $}   \Comment{$n$ is a passage at time step $t+1$}
            \If {SIMILARITY($m$, $n$) $>$ $n_{max}$}
                \State $n_{max}$  $\gets$ SIMILARITY($m$, $n$)
                \State $n_{id}$ $\gets$ $n$'s ID
            \EndIf
        \EndFor
        \If {$n_{max}$ $>$ $threshold$}
            \State $m_{id}$ $\gets$ $n_{id}$ \Comment{Update $m$'s ID so it is now identified as the same passage}
        \EndIf
    \EndFor
\EndFor
\EndProcedure
\end{algorithmic}
\end{algorithm}

\textbf{Passage-to-passage similarity using $n$-grams.} The similarity between a pair of passages is calculated using the normalized dot product of the $n$-grams of the two passages. Such a purely statistical similarity calculation based on $n$-grams is language independent and works well when the lengths of the two passages are arbitrary \cite{damashek1995gauging}. Character-level n-grams are obtained with the $n$ set to $5$. Mathematically, if $p$ and $q$ are two passages being compared, where $p_i$ is an $n$-gram in $p$ and $q_i$ is an $n$-gram in $q$, the first step is to calculate the weight of each distinct $n$-gram. In a passage $m$ (either $p$ or $q$), if $m_i$ is the number of occurrences of a distinct $n$-gram $i$ among J distinct n-grams, the weight of $i$ can be calculated as,

\begin{equation}
    x_i = \frac{m_i}{\sum_{j=1}^{J}m_j},
\end{equation}
where 
\begin{equation}
\sum_{j=1}^{J}x_{j} = 1.
\end{equation}
Then, for the two passages $p$ and $q$, if $x_{pj}$ is the relative frequency with which the $n$-gram $j$ occurs in passage $p$ and $x_{qj}$ is the relative frequency with which the $n$-gram $j$ occurs in passage $q$ (out of all possible $J$ $n$-grams in $p$ and $q$), the normalized dot product can be calculated as,
\begin{equation}
SIMILARITY(p,q) = \frac{\sum_{j=1}^{J}{x_{pj}x_{qj}}}{ \sqrt{\sum_{j=1}^{J}{x_{pj}^{2} \sum_{j=1}^{J}{x_{qj}^{2}}}}},
\end{equation}

\textbf{Preprocessing code data.} Transforming the revision history data of code data (e.g., a Python program), into a format that can be used to develop visualizations has a different set of challenges. Unlike natural language text, code lines can have high character similarities and yet have considerably different meanings. For instance, in code, the only difference between an inner for loop line and an outer for loop line serving different purposes can be two characters, $i$ and $j$. Because of this small coding difference, performing an all-to-all line comparison between two versions of code is ineffective. Consequently, the approach that was used to split natural language text paragraphs into sentences cannot be applied to split a code into lines. Instead, given two versions of the unsplit code, a `difference' calculating algorithm is applied to find common subsequences or shortest edits \cite{myers1986nd} on the entire code blocks. The central idea in calculating this `difference' is to formulate this problem of finding the longest common subsequence (LCS) and shortest edit script (SES) (known as the LCS/SES problem) as an instance of a single-source shortest path problem. The `npm diff' library implementation is used to obtain these code segments each with a specific label, `common,' `added,' or `removed.' These differences are obtained by considering each line as a unit, instead of considering each character or word as a unit. Following the idea similar to the one applied to natural language text versions, to obtain connections between the lines of code into adjacent versions of code, we initially generate random identities for all lines in all versions of the codes and update the identities of the sentences traversing backward from the most recent version of the code.

\textbf{Parameter-driven (re-)rendering of visualizations for analysis.} Interactive visualizations are rendered based on user-supplied values for parameters. This includes parameters such as $n$-gram size for calculating passage-to-passage similarity and the similarity threshold for determining the identities of passages (or lines in the case of code). Since all natural languages may not have space as the word separating character, \textit{user-defined} word delimiting character(s) and sentence delimiting character(s) make the tool accessible across a wide variety of languages.

% ToDo: Add URLs
\textbf{Technologies used in the web application.} The single-page \href{https://react.dev/}{React} web application was developed using several recent technologies. \href{https://tailwindcss.com/}{Tailwind} and JavaScript were used for building the website. The site includes \href{https://ckeditor.com/}{CKEditor} for allowing users to type text,
%\href{https://onecompiler.com/}{\hl{OneCompiler}}, 
and \href{https://judge0.com/}{Judge0} API \cite{hermann9245310}, an open-source online code execution system, for allowing users to code, \href{https://microsoft.github.io/monaco-editor/}{Monaco} text editor and diff viewer for typing and displaying code, and \href{https://github.com/plotly/react-plotly.js/}{Plotly} for displaying visualizations. The web application is hosted in \href{https://www.cloudflare.com/}{Cloudflare}. A user's process data can either be downloaded and stored for the user's record only or saved to Cloudflare's R2 buckets. This data is encrypted using the Advanced Encryption Standard (AES) implemented in the \href{https://www.npmjs.com/package/crypto-js}{`npm crypto-js'} library, with a user-supplied passcode, and can only be accessed with the passcode to decrypt it. This effectively serves as authentication for accessing the data. Data saved on the cloud R2 buckets are additionally encrypted using Cloudflare's encryption-at-rest (EAR) and encryption-in-transit (EIT) technologies.

\section{Results}

\begin{figure}[ht!]
\begin{center}
\includegraphics[width=0.8\columnwidth]{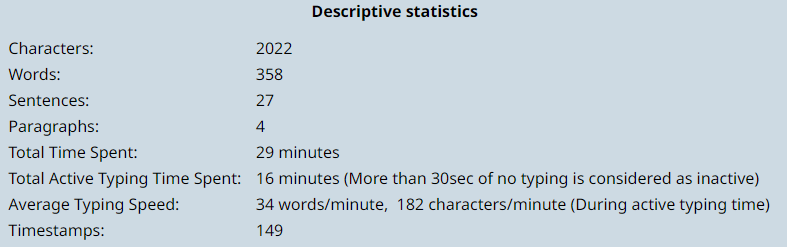}
\end{center}
\caption{An example table of \textbf{descriptive statistics}.}
\label{pstatistics}
\end{figure}

\begin{figure}[ht!]
\begin{center}
\includegraphics[width=1\columnwidth]{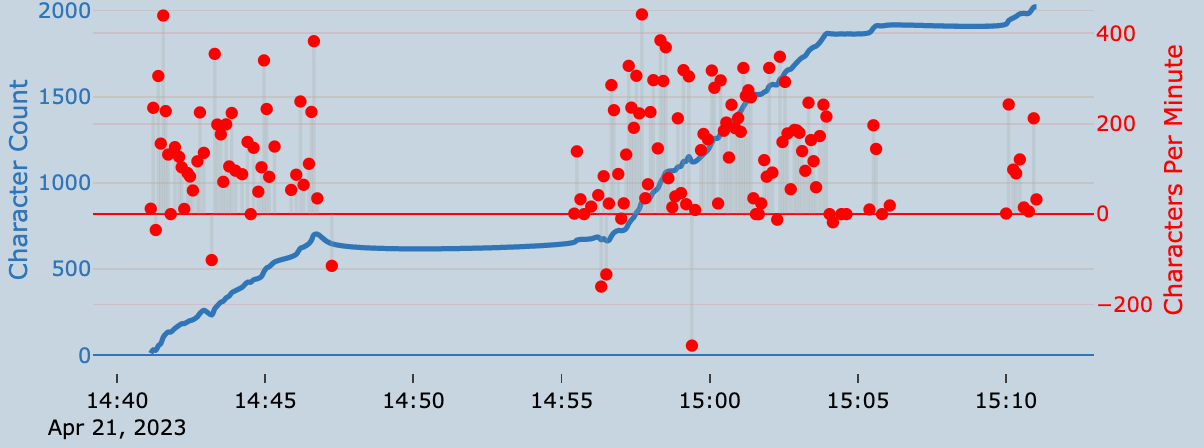}
\end{center}
\caption{An example dual-axis scatter and line plot displaying an increase in the \textbf{total characters typed} over time steps (blue line) and \textbf{characters per minute} at any time step (red points).}
\label{typing-speed}
\end{figure}

The revision history data collected from a user can be cleaned, processed, and summarized to generate informative summary tables and interactive visualizations. Descriptive statistics, such as the total number of typed characters, words, sentences, lines, paragraphs, total and active time spent on typing, and average typing speed, can be presented as summary sentences or a table (see Figure \ref{pstatistics}). These quantitative summaries can be accompanied by data graphics displaying the actual data points (as shown in Figure \ref{typing-speed}). Similarly, the visualizations of the process data can facilitate drilling into the revisions; they also provide an in-depth analysis of the overall process followed by a writer or a programmer. The next section discusses several interactive and non-interactive Process Visualizations (PVs).

\begin{figure}[ht!]
\begin{center}
\includegraphics[width=0.9\columnwidth]{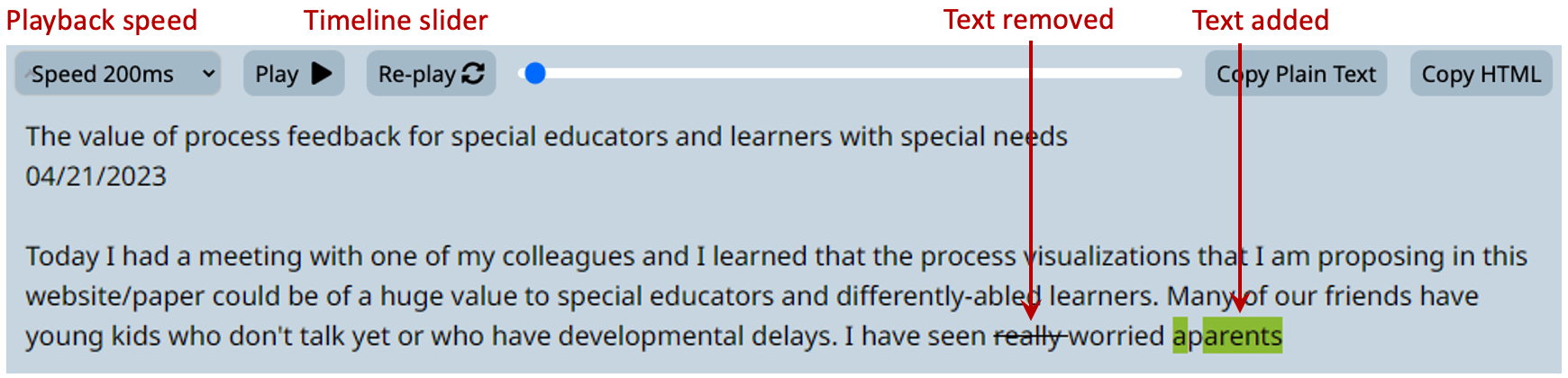}
\end{center}
\caption{An example interface for \textbf{playing 
 back} a writer's writing process. The green-highlighted copy shows text added at the time step and strikethrough formatting corresponds to text removed. The slider allows a user to control and view at any pace.}
\label{process-playback}
\end{figure}

\textbf{PV1: Process playback} (Figure \ref{process-playback})\textbf{.} A highly accessible and easy-to-understand technique of visualizing a learner's process is to play back their typing. A short (e.g., 30 seconds) playback can display the typing steps that a writer or a programmer took. Options to pause, stop, and replay at a user-controlled speed can make such a visualization interactive. During playback, at each time step, trails of text deleted (if any) and the trails of text added can be highlighted so a viewer can track where the learner was typing at the timestep. Such a process playback visualization may not provide any quantitative information but can be an engaging initial step toward further analysis.

\begin{figure}[ht!]
\begin{center}
\includegraphics[width=1\columnwidth]{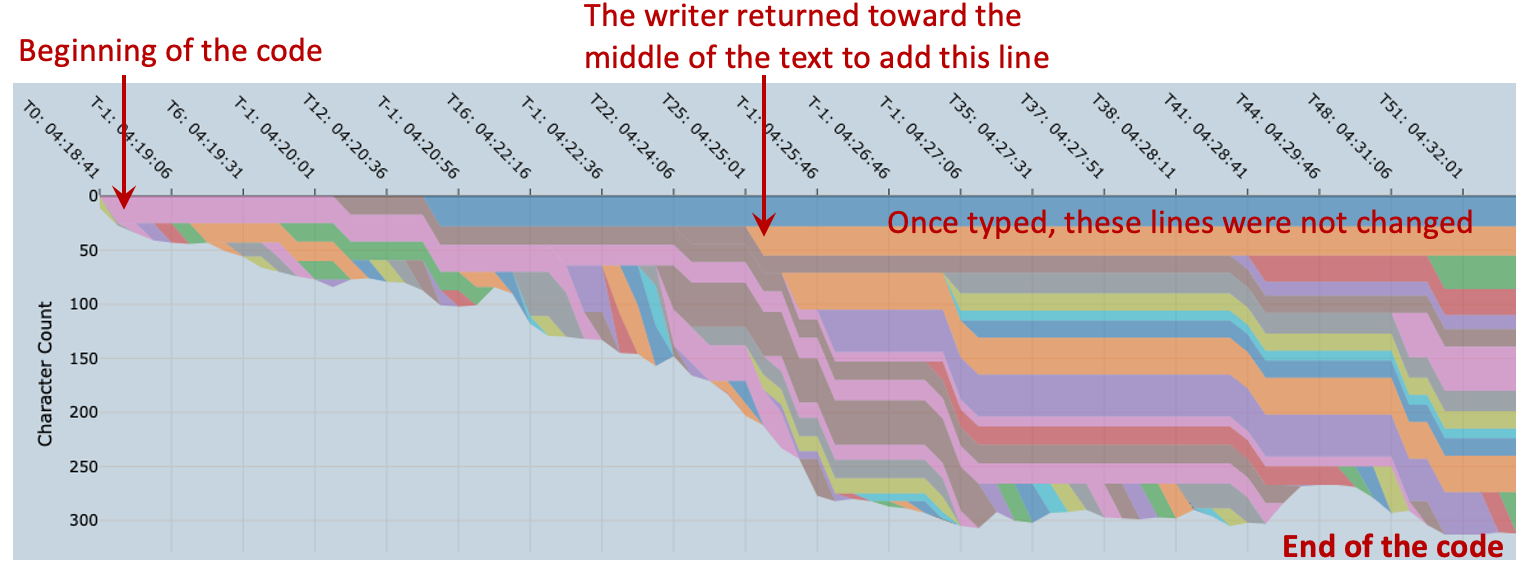}
\end{center}
\caption{An example stacked area plot displaying \textbf{sentence changes over time}. This graph displays at what time step all the sentences originated.}
\label{stacked-area}
\end{figure}

\textbf{PV2: Paragraph/sentence/line changes over time} (Figure \ref{stacked-area})\textbf{.} When analyzing one's writing or a programming process, it can be of interest to learn when a particular passage, paragraph, or sentence, in the case of writing and line in the case of code, originated or was removed. For instance, a sentence written at the beginning of the first paragraph may be pushed down throughout most of the writing time, only to later expand as a full paragraph. To visualize such process details at a passage level of interest, \textit{stacked area plots} may be used with unique colors assigned to each passage \cite{ghosh2012learning}. Such an area plot can also show the relative size of the sentences or paragraphs and how they change over time. Additionally, these stacked area plots can be interactive, i.e., a viewer can see the actual text of the passage at any given time step by hovering over the stacks.

\begin{figure}[ht!]
\begin{center}
\includegraphics[width=1\columnwidth]{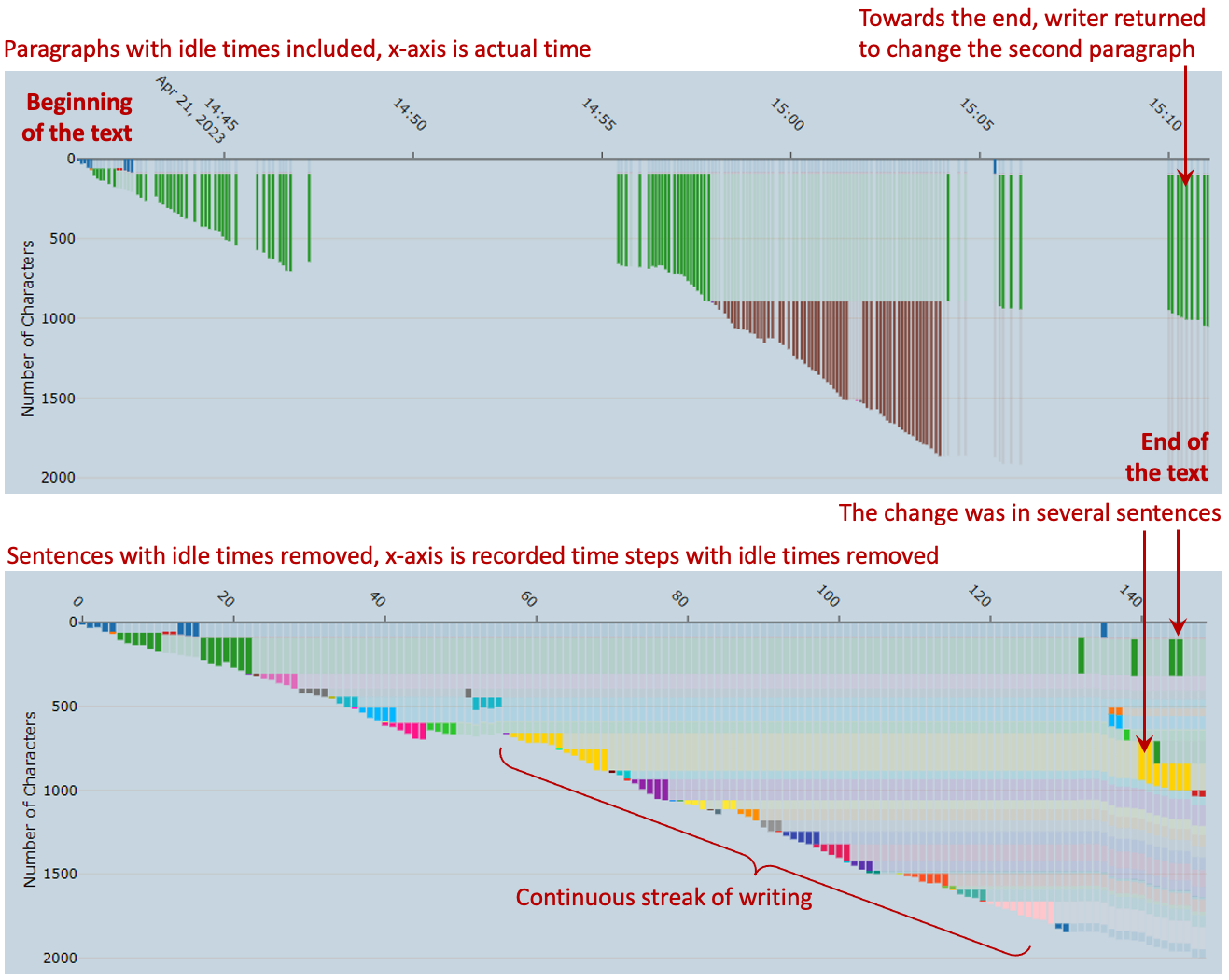}
\end{center}
\caption{Example stacked bar diagrams \textbf{highlighting active paragraph(s) (top figure) and active sentence(s) (bottom figure) at each time step}. Each color corresponds to a paragraph in the top figure and a sentence in the bottom figure. Idle times are ignored in the timeline. Hovering over any of the stacks displays the actual paragraph or sentence at the time step. Highlighted stacks represent the user's paragraph and sentence changes at the corresponding time step.}
\label{stacked-bar}
\end{figure}

\textbf{PV3: Highlight active paragraph/sentence/line at each timestep} (Figure \ref{stacked-bar})\textbf{.} An interactive stacked bar diagram can display each passage as a stack with time marked by x-axis. Such a detailed visualization can pinpoint (highlight) which passage a learner was working on at a specific time step during the writing or coding process. This visualization can also show how often a learner went back and revised a previously typed text. This can be achieved by highlighting only the passage being edited at that time step. 
An example application occurs when a learner is working on a revision task (starting with a bulk of text). In this case, the learner is looking at a stacked bar diagram to see if some paragraphs remain untouched. 

\begin{figure}[ht!]
\begin{center}
\includegraphics[width=0.65\columnwidth]{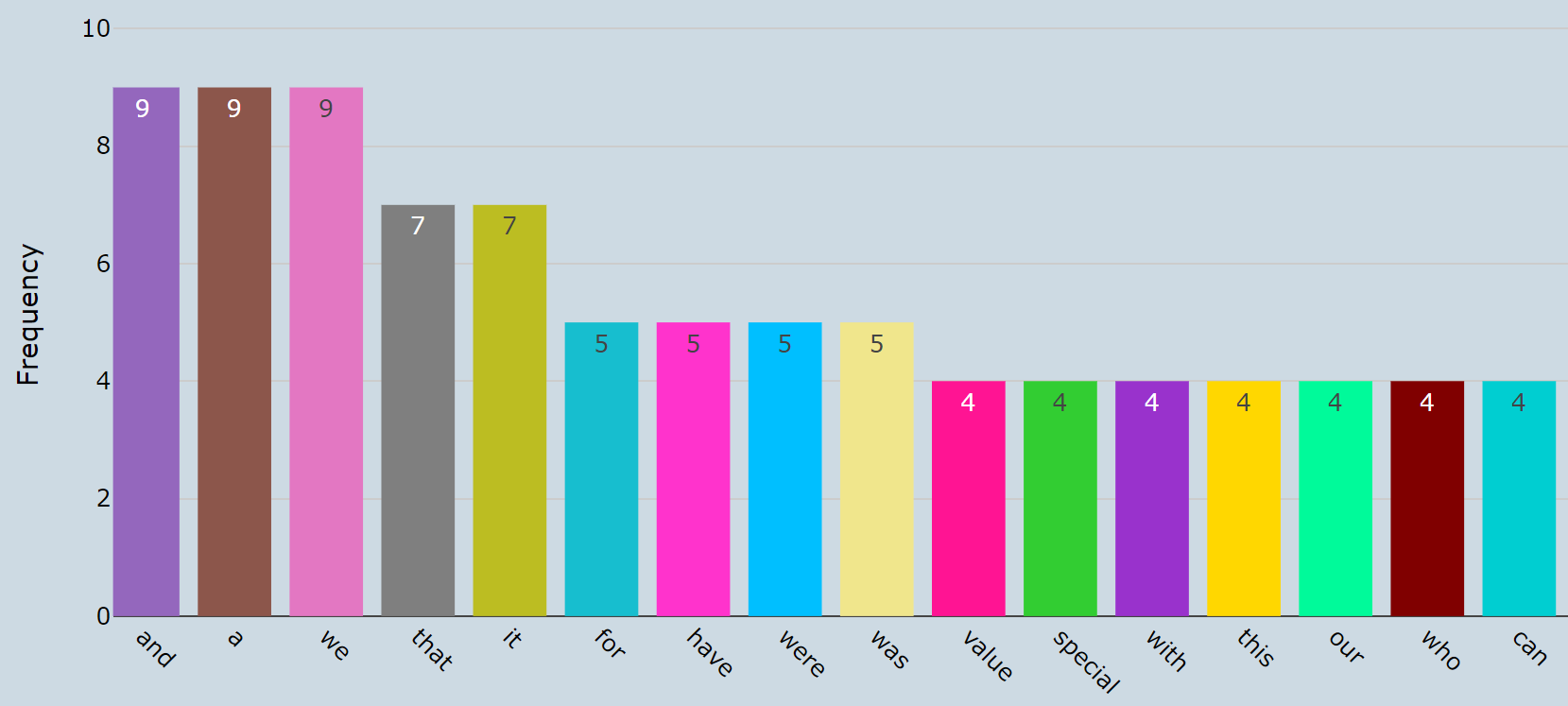}
\end{center}
\caption{An example bar diagram displaying \textbf{word frequencies}.}
\label{word-frequency}
\end{figure}

\textbf{PV4: Word frequencies} (Figure \ref{word-frequency})\textbf{.} In writing analysis, observing the frequencies of words typed by a learner can be informative. For instance, self-directed learners may be interested in observing if they overuse the article, \textit{the}, the adverbs \textit{very} and \textit{respectively} or conjunctive adverbs like \textit{however} and \textit{consequently}. Such observations may motivate them to increase their vocabulary and check the accuracy and necessity of overused and misused words. Although word frequency data can be calculated from the final text independent of the process, the frequency of words removed can also be informative. One caveat of counting words only based on user-defined character(s) but without the knowledge of a language is that these counts can be inaccurate for languages such as English where the same letters can be in lower or upper case. For instance, in such language-independent word frequency analysis, ``the" and ``The" would be identified as separate words. Word frequency can be plotted as standard bar diagrams displaying a certain percentage or a certain number of most used words and their corresponding usage frequency throughout the document.

\begin{figure}[ht!]
\begin{center}
\includegraphics[width=1\columnwidth]{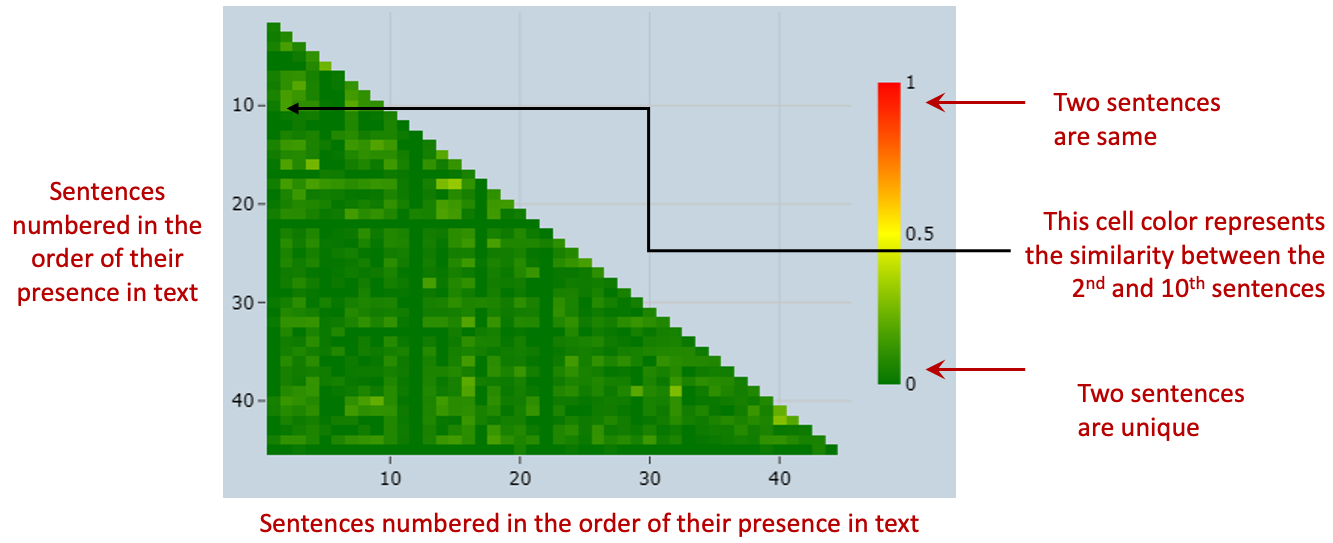}
\end{center}
\caption{An example heatmap plot for displaying \textbf{pairwise passage-to-passage similarity}. Both x-axis and y-axis represent the sentences shown in the order in which they appear in the final writing. Hovering over any cell of the heatmap shows the two sentences and their similarity score.}
\label{pairwise-similarity}
\end{figure}

\textbf{PV5: Passage-to-passage pairwise similarities} (Figure \ref{pairwise-similarity})\textbf{.} Writers can also fall into the habit of repeatedly using the same phrases that fail to inform or can be deleted, such as \textit{it should be noted that} or \textit{it is well known that}. Visualizing pairwise similarity between all sentences in the final version of the text can help identify such habits. A visualization that serves this purpose is an interactive heatmap where each cell represents a similarity score between pairs of sentences. Displaying the actual sentence pairs by allowing them to hover over the cells can add interactivity.

\textbf{PV6: Words/characters change over time} (Figure \ref{typing-speed})\textbf{.} Changes in the number of words or characters over time can be visualized as a line diagram for cumulative and as a lollipop chart (or a bubble chart) for non-cumulative review. In a bubble chart, the bubble sizes can represent the number of characters added or removed. One approach to display bubbles at a consistent location in the y-axis is to set the y-axis based on the size of the change. Since changes can have a large variation data can be log-normalized.

\begin{figure}[ht!]
\begin{center}
\includegraphics[width=1\columnwidth]{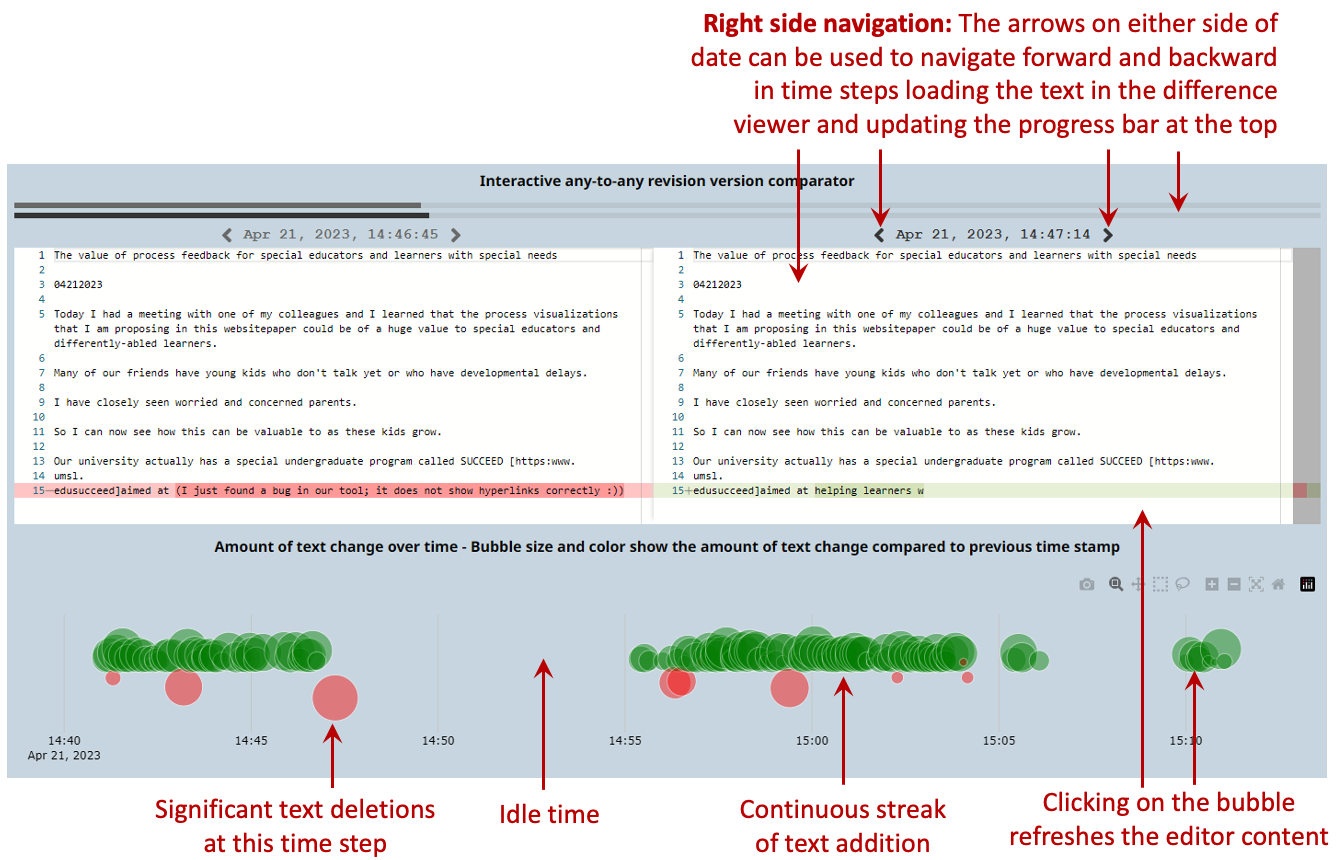}
\end{center}
\caption{An example \textbf{interactive any-to-any revision version comparator} consisting of three components: 1) two date navigators at the top with arrows on either side of the dates, with progress bars on top connected to the dates (top progress bar for left date and bottom progress bar for right date); 2) a difference viewer shows two versions of text dictated by either the clicking on the date arrows or clicking on the bubble chart points at the figure's bottom; 3) a bubble chart shows the timeline where the bubble sizes correspond to the number of characters added or removed. Panning and zooming on the bubble chart facilitates easy micro/macro analysis. Clicking on any bubble loads the changes made in that particular version in the difference viewer windows. Green bubbles indicate that the version has text added and red indicates text was deleted. Also, hovering over the bubbles shows total characters added/removed.}
\label{version-comparator}
\end{figure}

\textbf{PV7: Interactive any-to-any revision version comparator} (Figure \ref{version-comparator})\textbf{.} Users interested in an in-depth analysis will want to observe the text added or removed between any two time steps in the revision history. Users also need an interactive visualization tool that enables them to browse the entire timeline and select a range for viewing the changes. To do so involves effectively selecting two time steps. This can be achieved by combining three tools: 1) a difference viewer (i.e., the Monaco diff viewer), 2) a timeline chart (i.e., a bubble chart) which can show all the time steps, and 3) a date-based navigator. Users can then interact through the bubble chart or navigate directly through the diff viewer. Such an interactive visualization can facilitate in-depth analysis, allowing the user to ask why a sentence or phrase was removed at a certain time.

\begin{figure}[ht!]
\begin{center}
\includegraphics[width=1.0\columnwidth]{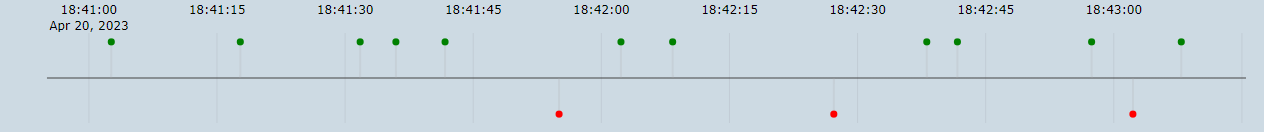}
\vspace{5pt}
\caption{An example timeline showing \textbf{successful executions} (green points) and \textbf{executions with errors} (red points) in a computer programmer's coding process.}
\label{code-execution}
\end{center}
\end{figure}

\textbf{PV8: Timeline showing successful and unsuccessful code executions} (Figure \ref{code-execution})\textbf{.} In the case of programming tasks, displaying a timeline plot with points indicating the executions of code, highlighting both successful and unsuccessful executions, can help users review how often they executed their code.

\section{Discussion}

\subsection{Exploratory data visualizations as clarifying tools} 

Data graphics are widely used as an explanation tool. In a more general, all-inclusive context, data graphics can serve the purpose of explanation, exploration, or both. While explanatory data graphics can communicate a point or display a pattern or a concept, exploratory data visualizations invite the viewer to discover information \cite{frankel2012visual}. While discussing the fundamental principles of analytical design, in his book ``Beautiful Evidence” \cite{tufte2006beautiful}, Edward Tufte lists ``completely integrating words, numbers, images, and diagrams” as one of the core principles. Although these exploratory data visualizations can appear complex at first, they can be vehicles for new insights. Alberto Cairo in Truthful Art \cite{cairo2016truthful}, noted insightfulness as one of the most important qualities in his presentation of \textit{five qualities} of great visualizations. Adding interactivity to visualizations can strengthen this integration. Interactive and integrative visualizations can facilitate two fundamental acts of science: description and comparison. Several aspects of interactive data visualization can, however, make them inaccessible. For increased public accessibility, these visualizations should be improved for user-friendliness and high usability. 

\subsection{Acquiring feedback from insightful Process Visualizations} 

The Process Visualizations clarify the concepts and steps of the process. This provides tools for acquiring feedback---from self, a peer, or an educator---on several dimensions of a writer's or programmer's process. For instance, if a writer is focused on improving their prewriting skills, these visualizations can enable exploring how frequently the writer switches to other modes, such as revising and editing (see Figure \ref{stacked-bar}). While it may be recommended to finish prewriting and then focus on revision, it is possible to polish paragraphs while writing. Prewriting embodies the initial steps of writing. A writer's approach is usually an iterative process; thus, it can be insightful to learn what process a writer follows to finalize an initial prewrite. Similarly, in learning and practicing the `test first' approach to programming, Process Visualizations can enable exploration of how often the programmer revises the test cases as they start to code the actual solution and how often they revise the solution as the test cases evolve. In general, learners could also use these visualizations for self-revelation, direction, and improvement, and the visualizations can provide immediate answers to questions such as: (1) where in the process did they spend most of the time, (2) how much time did they spend on creating the first draft vs. revising and editing the draft, (3) which paragraphs were edited and revised, and (4) which paragraphs were not edited. Overall, the visualizations facilitate exploring, looking back, and looking into one's process and approach. Furthermore, learners can engage in \textit{inquiry} by focusing on the process, regardless of whether the processing was originated by the learner, a peer, or an educator. Exploring and understanding the theories behind teaching and learning followed by the engagement of all parties toward making the most of every learning experience is the portal to learning through communication, feedback, improvement, and success. Engaging in inquiry and agreeing on the \textit{theories that drive teaching and learning} are powerful. Whitney \cite{whitney2008beyond} summarized theorizing and engagement by saying:

\begin{quote}
``Engaging in inquiry means not only learning practices recommended by others or perfecting the practical execution of a set of teaching strategies but, rather, theorizing about teaching and learning in a way that then frames future interpretation and decision-making.'' (Beyond strategies: Teacher practice, writing process, and the influence of inquiry, 2008, p. 205).
\end{quote}

%Self-Direction Plan: Make sure you understand what the instructor is asking you to do. What is the overall goal of the project? What is the timeline? How can I budget my time to make sure I finish all tasks on time? 
%Self-Monitoring: To self-monitor, determine the strategy that works best for you. Honestly prepare a list of how this assignment will challenge your strengths and weaknesses. Look for new resources to better explain your unanswered question. Find out why you previously failed to master a skill you need and determine to find solutions so you will not  fail again. 
%Self-evaluating: To self-evaluate, ask: How successful was I in completing this task? Did I complete all the tasks? If not, why? What can I do to improve my timing, performance, and results? What kept me from being more efficient? Did I manage my time well? What do I need to do differently if I am asked to complete an assignment like this again?

\subsection{Securing personal process data} 

Individual creations such as a piece of writing, code, or art are intellectual properties, and their security is paramount in the contemporary digital landscape. The process data from an individual (i.e., the revision history) can be at least as important as the outcome. However, process data tends to be much more sensitive and personal than the final outcome. Such data could be tampered with or exchanged with malicious intent. For instance, if trained on process data, deep learning methods could potentially generate ``deep process fakes," which can have disastrous and unintended consequences. The deep fake scenarios \cite{chesney2019deep} have brought out the ability to alter photos and videos to the point of misrepresenting a person's political stance, corporate identity, or involvement in a crime. Hence, securing and sharing personal process data with only trusted parties is critical. It is also urgent to develop strict policies and restrictions for machine or deep learning algorithms that are applied to individual process data. The current version of the Process Visualization techniques proposed in this work, including the n-gram similarity calculations, does not use any form of machine learning or deep learning.

\subsection{Are learners monitored or controlled during their creative process?}

The current release of our web application does not save any user data to our online servers, except at required times such as the time of loading the single-page web application, downloading a task description (question), explicitly saving a question or a response to the online storage server, verifying a human user, or when executing a computer code (to compile). However, any third-party components used could be using their own data collection techniques. All revision histories of a user's data are saved locally in the user's Internet browser's local storage. In other words, users have complete control over what process data they want to share and when. For instance, the web application does not track if a learner working on a project clears history, starts over,  or abandons the writing process altogether. 

The web application can be used entirely as a self-improvement tool. If learners wish to share their process, they can self-review it first and then decide if they want to share it. The current version of the application does not have any authentication built-in (e.g., logging in or signing in). The intent here is to make the website highly accessible to users since requiring an email address and forgetting passwords won't be a concern. Hence, users are not required to include any private information. \textit{If a user writes their content, views the Process Visualizations, and downloads the data without saving it to our servers, our application has no knowledge of any such process and does not attempt to track them.} In sum, users have a clear choice to discontinue revision logging at any time, and if they do decide to share, can review their process data first on their own.

If the technique of maintaining a revision history was to be used in non-transparent ways (i.e., in exams, tests, or standardized language/programming tests), learners could feel vulnerable, monitored, and even controlled. This type of non-transparency could cause learners to resist using any education/learning system if given a choice, including the system presented herein. As an alternative, in a setting where learners feel monitored or controlled, an educator can motivate their learners by showing (modeling) their creative process using visualizations. For instance, an educator teaching writing can demonstrate how common it is to delete text during a revision process and an educator teaching programming could demonstrate how common it is to run into errors and spend time debugging.

\subsection{Will educators' assessment time increase?}

One concern that may be raised when exploring and interpreting Process Visualizations is that educators' assessment time could increase. It can be argued that educators worldwide are already struggling and need more time to provide feedback that promotes learning. Moreover, educators are increasingly concerned about how to build students’ self-efficacy \cite{nicol2006formative}. Unfortunately, finding time to develop tools that can help build students’ self-efficacy and promote learning is a hurdle not many educators can achieve. Teachers and students only have so many hours each day, which is usually strictly regimented. Additionally, teachers cannot possibly know all the individual problems students have had in their learning process. At times, not even the student knows the problems that are causing barriers to learning. How, then, will educators be able to dig into the process of learning and providing feedback on the process? This is a genuine initial concern; however, several premises must be considered. 

First, compared to reading an outcome (essay or code), visualizations can be faster to interpret, particularly when the structure of the visualization is consistent, with only the data changing. The first few exploratory activities and interactions will take time. However, once the structure and features of the visualizations are learned, the focus will naturally shift to seeing the story being unveiled by the data. These visualizations, however, should be improved for the highest accessibility and user-friendliness.

Second, after exploring a few varieties that may be possible within a visualization structure, users will develop knowledge about the range of possible variations and gradually become efficient at interpreting meanings from the visualizations. Short and dedicated tutorial videos on each visualization, aimed at enhancing visualization literacy, can also help the users learn how to interact with the visualizations.

Third, not all learners and educators should use the visualization toolbox for every task. Learners may only use it occasionally to learn about their style or with their peers to learn from each other. Similarly, educators may use it only for a specific task or for a learner who would benefit from such feedback. 

Fourth, in situations where the outcome is more important than the process, learners and educators can discard the Process Visualization approach altogether because these visualizations do not replace the outcome; they only supplement it. The premise of this research is that if learners and educators appropriately divide their time between reading the completed writing or coding tasks and exploring the Process Visualizations (i.e., consider both: the outcome and the process), the two can complement each other and together serve as a powerful resource to either self-learn or acquire feedback from others.

\section{Conclusion}

This work introduced several Process Visualizations aimed at facilitating writers and programmers to acquire feedback on their processes. 
The visualizations can encourage learners and educators to focus on the process and serve as tools to analyze their creative and cognitive processes.
These visualizations and ideas can also serve as an inspiration for designing similar tools for learners and practitioners outside of writing and coding. For example, designers, music composers, artists, filmmakers, and architects could benefit from similar process-focused visualizations and analysis. 
%Facilitating a creative unhindered focus on learning results in a more balanced conception of any topic. It also leads to a more creative approach to solving problems. Overcoming the fear of learning in areas that were previously difficult (and too scary to approach for some) can create new groups of researchers whose skills are balanced and fearless.
Facilitating a focus on creative endeavor processes can also strengthen the core of current education systems, allowing learners to thrive.
Such a process elucidating tool can also be helpful for special educators to collect data and write  goals in the individualized education plans (IEPs) for differently-abled learners.
Meanwhile, Process Visualizations and feedback may emerge as a new interdisciplinary field at the crossroads of psychology, education, computer science, and data visualization and can be quickly adopted in several existing systems and institutions throughout the future. For this to be effective and successful, the security of personal process data is also paramount.

\vspace{10pt}

\newpage

\textbf{\large{Availability}}

The Process Feedback website is ready to be tested by educators and learners and is publicly available to all at 
\href{https://www.processfeedback.org}{www.processfeedback.org}.

\vspace{10pt}

\textbf{\large{Acknowledgments}}

I am deeply grateful to the individuals who have contributed to the development of this manuscript. Their expertise and thoughtful insights have been essential in shaping and refining the content. Specifically, I extend my heartfelt appreciation to Dr. Shea Kerkhoff, Dr. Abderrahmen Mtibaa, Dr. Sambriddhi Mainali, Jason Wagstaff, and graduate students Shaney Flores and Kate Arendes at the University of Missouri-St. Louis, as well as undergraduate student Nilima Kafle, for engaging in several stimulating discussions and for dedicating their time to providing insightful feedback on the manuscript. Furthermore, I would like to acknowledge Dr. Manu Bhandari and Dr. Khem Aryal at Arkansas State University, Dr. Jie Hou from Saint Louis University, as well as Bishal Shrestha and Nitesh Kafle from Kathmandu, Nepal, for their valuable discussions and comments on the manuscript. I am also grateful for the recommendations provided by Dr. Amber Burgett at the University of Missouri-St. Louis and Dr. Laura March at the University of Missouri-Columbia. Finally, I am grateful to Carla Roberts at Preferred Copy Editing for editing the manuscript at a short notice.

\vspace{10pt}

% Resource: https://shantoroy.com/latex/how-to-write-algorithm-in-latex/
%\renewcommand{\thealgorithm}{} % For removing numbering
%\renewcommand{\algorithmicrequire}{\textbf{Inputs:}}
%\renewcommand{\algorithmicensure}{\textbf{Outputs:}}
%\algnewcommand\algorithmicforeach{\textbf{for each}}
% \algdef{S}[FOR]{ForEach}[1]{\algorithmicforeach\ #1\ \algorithmicdo}
% \begin{algorithm}
% \small
% \caption{Calculate similarity between two text passages $P$ and $Q$}
% \label{myalgorithm2}
% \begin{algorithmic}[1]
% \Require~~
% \Statex {$n$, `n' in n-grams} \Comment{2 as default value}
% \Statex {$s$, n-gram scope (word or character level)} \Comment{Character-level is slower but more effective}
% \Statex {$w$, character/s serving as `end of a word' marker} \Comment{`space' character as default value}
% \Ensure~~
% \Statex {$s$, similarity between the passages $P$ and $Q$}
% \Procedure{SIMILARITY}{}
% \State $P_{n-grams}$ $\gets$ NGRAMS($P$, $n$, $s$, $w$)   \Comment{Calculates n-grams for passage $P$}
% \State $Q_{n-grams}$ $\gets$ NGRAMS($Q$, $n$, $s$, $w$) \Comment{Calculates n-grams for passage $Q$}
% \State $Intersection$ $\gets$ count of n-grams common in $P_{n-grams}$ and $Q_{n-grams}$  
% \State $Union$ $\gets$ count of union n-grams in $P_{n-grams}$ and $Q_{n-grams}$  
% \If {$Union$ $=$ $0$}
%     \State return 0
% \EndIf
% \State return $Union$ / $Intersection$ \Comment{Similarity score between 0 and 1}
% \EndProcedure
% \end{algorithmic}
% \end{algorithm}

\bibliography{references}

\begin{thebibliography}{10}

\bibitem{langer1987writing}
Judith~A Langer and Arthur~N Applebee.
\newblock {\em How Writing Shapes Thinking: A Study of Teaching and Learning.
  NCTE Research Report No. 22.}
\newblock ERIC, 1987.

\bibitem{klein2016trends}
Perry~D Klein and Pietro Boscolo.
\newblock Trends in research on writing as a learning activity.
\newblock {\em Journal of writing research}, 7(3):311--350, 2016.

\bibitem{whitney2008beyond}
Anne Whitney, Sheridan Blau, Alison Bright, Rosemary Cabe, Tim Dewar, Jason
  Levin, Roseanne Macias, and Paul Rogers.
\newblock Beyond strategies: Teacher practice, writing process, and the
  influence of inquiry.
\newblock {\em English Education}, 40(3):201--230, 2008.

\bibitem{emig1977writing}
Janet Emig.
\newblock Writing as a mode of learning.
\newblock {\em College composition and communication}, 28(2):122--128, 1977.

\bibitem{pea1984cognitive}
Roy~D Pea and D~Midian Kurland.
\newblock On the cognitive effects of learning computer programming.
\newblock {\em New ideas in psychology}, 2(2):137--168, 1984.

\bibitem{kurland1986study}
D~Midian Kurland, Roy~D Pea, Catherine Clement, and Ronald Mawby.
\newblock A study of the development of programming ability and thinking skills
  in high school students.
\newblock {\em Journal of Educational Computing Research}, 2(4):429--458, 1986.

\bibitem{popat2019learning}
Shahira Popat and Louise Starkey.
\newblock Learning to code or coding to learn? a systematic review.
\newblock {\em Computers \& Education}, 128:365--376, 2019.

\bibitem{flavell1979metacognition}
John~H Flavell.
\newblock Metacognition and cognitive monitoring: A new area of
  cognitive--developmental inquiry.
\newblock {\em American psychologist}, 34(10):906, 1979.

\bibitem{georghiades2004general}
Petros Georghiades.
\newblock From the general to the situated: Three decades of metacognition.
\newblock {\em International journal of science education}, 26(3):365--383,
  2004.

\bibitem{national2000people}
National~Research Council et~al.
\newblock {\em How people learn: Brain, mind, experience, and school: Expanded
  edition}.
\newblock National Academies Press, 2000.

\bibitem{brame2019science}
Cynthia~J Brame.
\newblock {\em Science teaching essentials: short guides to good practice}.
\newblock Academic Press, 2019.

\bibitem{hancock2021supporting}
Emilie Hancock and Gulden Karakok.
\newblock Supporting the development of process-focused metacognition during
  problem-solving.
\newblock {\em PRIMUS}, 31(8):837--854, 2021.

\bibitem{baten2017relevance}
Elke Baten, Magda Praet, and Annemie Desoete.
\newblock The relevance and efficacy of metacognition for instructional design
  in the domain of mathematics.
\newblock {\em ZDM}, 49:613--623, 2017.

\bibitem{langer1986children}
Judith~A Langer.
\newblock {\em Children reading and writing: Structures and strategies.}
\newblock Ablex Publishing, 1986.

\bibitem{erdogmus2005effectiveness}
Hakan Erdogmus, Maurizio Morisio, and Marco Torchiano.
\newblock On the effectiveness of the test-first approach to programming.
\newblock {\em IEEE Transactions on software Engineering}, 31(3):226--237,
  2005.

\bibitem{hattie2007power}
John Hattie and Helen Timperley.
\newblock The power of feedback.
\newblock {\em Review of educational research}, 77(1):81--112, 2007.

\bibitem{lipnevich2009effects}
Anastasiya~A Lipnevich and Jeffrey~K Smith.
\newblock Effects of differential feedback on students’ examination
  performance.
\newblock {\em Journal of Experimental Psychology: Applied}, 15(4):319, 2009.

\bibitem{xiao2010effects}
Yun Xiao.
\newblock The effects of training in peer assessment on university students'
  writing performance and peer assessment quality in an online environment.
\newblock {\em Doctor of Philosophy (PhD), Dissertation, Teaching and Learning,
  Old Dominion University}, 2010.

\bibitem{cauley2010formative}
Kathleen~M Cauley and James~H McMillan.
\newblock Formative assessment techniques to support student motivation and
  achievement.
\newblock {\em The clearing house: A journal of educational strategies, issues
  and ideas}, 83(1):1--6, 2010.

\bibitem{mcmillan2008student}
James~H McMillan and Jessica Hearn.
\newblock Student self-assessment: The key to stronger student motivation and
  higher achievement.
\newblock {\em Educational horizons}, 87(1):40--49, 2008.

\bibitem{paulson2017impact}
Kathy Paulson~Gjerde, Margaret~Y Padgett, and Deborah Skinner.
\newblock The impact of process vs. outcome feedback on student performance and
  perceptions.
\newblock {\em Journal of Learning in Higher Education}, 13(1):73--82, 2017.

\bibitem{black1998assessment}
Paul Black and Dylan Wiliam.
\newblock Assessment and classroom learning.
\newblock {\em Assessment in Education: principles, policy \& practice},
  5(1):7--74, 1998.

\bibitem{shepard2000role}
Lorrie~A Shepard.
\newblock The role of assessment in a learning culture.
\newblock {\em Educational researcher}, 29(7):4--14, 2000.

\bibitem{yorke2003formative}
Mantz Yorke.
\newblock Formative assessment in higher education: Moves towards theory and
  the enhancement of pedagogic practice.
\newblock {\em Higher education}, 45:477--501, 2003.

\bibitem{dixson2016formative}
Dante~D Dixson and Frank~C Worrell.
\newblock Formative and summative assessment in the classroom.
\newblock {\em Theory into practice}, 55(2):153--159, 2016.

\bibitem{boud1995assessment}
David Boud.
\newblock Assessment and learning: contradictory or complementary.
\newblock {\em Assessment for learning in higher education}, pages 35--48,
  1995.

\bibitem{damashek1995gauging}
Marc Damashek.
\newblock Gauging similarity with n-grams: Language-independent categorization
  of text.
\newblock {\em Science}, 267(5199):843--848, 1995.

\bibitem{myers1986nd}
Eugene~W Myers.
\newblock An o (nd) difference algorithm and its variations.
\newblock {\em Algorithmica}, 1(1-4):251--266, 1986.

\bibitem{hermann9245310}
Herman~Zvonimir Došilović and Igor Mekterović.
\newblock Robust and scalable online code execution system.
\newblock In {\em 2020 43rd International Convention on Information,
  Communication and Electronic Technology (MIPRO)}, pages 1627--1632, 2020.

\bibitem{ghosh2012learning}
Satrajit~S Ghosh, Arno Klein, Brian Avants, and K~Jarrod Millman.
\newblock Learning from open source software projects to improve scientific
  review.
\newblock {\em Frontiers in computational neuroscience}, 6:18, 2012.

\bibitem{frankel2012visual}
Felice Frankel and Angela~H DePace.
\newblock {\em Visual strategies: A practical guide to graphics for scientists
  \& engineers}.
\newblock Yale University Press, 2012.

\bibitem{tufte2006beautiful}
Edward~R Tufte.
\newblock {\em Beautiful evidence}, volume~1.
\newblock Graphics press Cheshire, CT, 2006.

\bibitem{cairo2016truthful}
Alberto Cairo.
\newblock {\em The truthful art: Data, charts, and maps for communication}.
\newblock New Riders, 2016.

\bibitem{chesney2019deep}
Bobby Chesney and Danielle Citron.
\newblock Deep fakes: A looming challenge for privacy, democracy, and national
  security.
\newblock {\em Calif. L. Rev.}, 107:1753, 2019.

\bibitem{nicol2006formative}
David~J Nicol and Debra Macfarlane-Dick.
\newblock Formative assessment and self-regulated learning: A model and seven
  principles of good feedback practice.
\newblock {\em Studies in higher education}, 31(2):199--218, 2006.

\end{thebibliography}
\bibliographystyle{unsrt}

\end{document}